%% file: paper.tex
\documentclass[letterpaper,twocolumn,10pt]{article}
\usepackage{usenix-2020-09}

\usepackage{xcolor}
\usepackage{enumerate}
\usepackage{hyperref}
\usepackage{xspace}
\usepackage{tlatex}
\usepackage{multirow}
\usepackage{graphicx}
\usepackage{siunitx}
\usepackage{booktabs}
\usepackage[font={it,small}]{caption,subcaption} %
\usepackage[compact]{titlesec}
\usepackage{listings}
\usepackage{enumitem}
\usepackage[normalem]{ulem}

\setboolean{shading}{true}

\usepackage[compact]{titlesec}
\titlespacing{\section}{0pt}{2ex}{1ex}
\titlespacing{\subsection}{0pt}{1ex}{1ex}
\titlespacing{\subsubsection}{0pt}{0.5ex}{0ex}
\usepackage[font=small]{caption}
\newcommand{\oom}[1]{{10\textsuperscript{\small #1}}}
\newcommand{\circlelabel}[1]{{\large \textcircled{\small #1}}}

\newif\ifcomments
\commentsfalse %

\ifcomments
    \newcommand{\hh}[1]{{\color{blue}HH: #1}}
    \newcommand{\ac}[1]{{\color{red}AC: #1}}
    \newcommand{\ea}[1]{{\color{brown}EA: #1}}
    \newcommand{\nc}[1]{{\color{orange}NC: #1}}
    \newcommand{\mk}[1]{{\color{purple}MK: #1}}
\else
    
    \newcommand{\hh}[1]{}
    \newcommand{\ac}[1]{}
    \newcommand{\ea}[1]{}
    \newcommand{\nc}[1]{}
    \newcommand{\mk}[1]{}
\fi

\newif\ifcompress
\compresstrue %
\ifcompress
  \setlist[itemize]{leftmargin=*}
  \setlist[enumerate, 1]{1.} %
  \setlist{itemsep=0pt,parsep=0pt}             %
  \usepackage[subtle,wordspacing=normal]{savetrees} %
  \titlespacing*{\section} {0pt}{4pt}{2pt}
  \titlespacing*{\subsection} {0pt}{2pt}{2pt}
\fi
\newcommand{\summary}[1]{}

\newcommand{\ghissue}[1]{\href{https://github.com/microsoft/CCF/issues/#1}{\##1}}

{}
\newcommand{\tlaplus}{TLA\textsuperscript{\textup{+}}\xspace}

\newtheorem{property}{Property}

\begin{document}
\title{Smart Casual Verification of the Confidential Consortium Framework}

\author{Heidi Howard$^*$, Markus A. Kuppe$^*$, Edward Ashton$^*$, Amaury Chamayou$^*$, Natacha Crooks$^*$$^\dagger$ \\
Azure Research, Microsoft$^*$ \hspace{2cm} UC Berkeley$^\dagger$}

\maketitle

\input{0-abstract}

\input{1-intro}
\input{2-ccf}

\input{3-tla}
\input{4-consensus}
\input{5-consistency}

\input{6-trace}

\input{7-results}
\input{8-lessons}
\input{9-related}

\input{10-conclusion}

\bibliographystyle{plainurl}
\bibliography{../refs}

\appendix

\end{document}

%% file: 0-abstract.tex
\begin{abstract}
	The Confidential Consortium Framework (CCF) is an open-source platform for developing trustworthy and reliable cloud applications.
	CCF powers Microsoft's Azure Confidential Ledger service and as such it is vital to build confidence in the correctness of CCF's design and implementation.
	This paper reports our experiences applying \emph{smart casual verification} to validate the correctness of CCF's novel distributed protocols, focusing on its unique distributed consensus protocol and its custom client consistency model.
	We use the term smart casual verification to describe our hybrid approach, which combines the rigor of formal specification and model checking with the pragmatism of automated testing, in our case binding the formal specification in \tlaplus{} to the C++ implementation.
	While traditional formal methods approaches require substantial buy-in and are often one-off efforts by domain experts, we have integrated our smart casual verification approach into CCF's CI pipeline, allowing contributors to continuously validate CCF as it evolves.
	We describe the challenges we faced in applying smart casual verification to a complex existing codebase and how we overcame them to find six subtle bugs in the design and implementation before they could impact production.
\end{abstract}

%% file: 1-intro.tex
\section{Introduction}\label{sec:intro}

\summary{CCF is important/novel/interesting}
The Confidential Consortium Framework (CCF)~\cite{Howard23,CCF} is a general-purpose platform for developing trustworthy and highly available cloud applications. 
CCF combines centralized compute with decentralized trust, supporting deployment on untrusted cloud infrastructure and transparent governance by mutually untrusted parties.
CCF achieves this by leveraging hardware-based trusted execution environments (TEEs) for remotely verifiable confidentiality and code integrity~\cite{Russinovich21,Russinovich24}, coupled with state machine replication backed by an auditable immutable ledger for data integrity and high availability.
CCF is trusted in production by services such as Azure Confidential Ledger~\cite{ACL}, a tamper-proof append-only ledger, which is utilized for storing critical data like digests for integrity-protection with SQL Server~\cite{Antonopoulos2020,SQLLedger} and Azure Immutable Blob Storage~\cite{ACL-blobs}.
Open source applications of CCF range from code transparency~\cite{Delignat23,scitt} and mediation of multi-party data sharing~\cite{PDO}, to decentralized identity~\cite{did-ccf}, privacy-preserving ad auctions~\cite{privacy-sandbox}, and confidential storage~\cite{Jeffery24,LSKV}.

This paper summarizes our experience with verifying the correctness guarantees of the distributed protocols in CCF, a large-scale, production distributed system, using \tlaplus~\cite{SpecifyingSystems,Lamport94}. We have five primary requirements that guided our approach to verification:

\textbf{(1) Verify high-level distributed safety properties.} 
We aim to check the correctness of our custom distributed consensus protocol, including its dynamic reconfiguration logic. CCF's consensus logic, though based on Raft~\cite{Ongaro14} has been sufficiently modified such that it is now based on an unproven algorithm. This is a common problem: the Chubby authors~\cite{Burrows06} similarly note that the additional requirements imposed by a real-world system over a vanilla Paxos~\cite{Lamport98} implementation require significant changes. Moreover, distributed consensus protocols are notoriously hard to get right, and even well-known protocols, including PBFT~\cite{Castro99}, Egalitarian Paxos~\cite{Moraru13}, and Zyzzyva~\cite{Kotla10}, as well as the previously mentioned Raft and Chubby, have been found to contain subtle bugs~\cite{Berger21,RaftConfigBug,Abraham2017,Amos15,whittaker21,Sutra20,Michael17}. 
The chosen verification strategy must therefore be able to check properties across distributed nodes, handling asynchrony, concurrency, and non-determinism in the order that events, such as timeouts and message delivery, occur as well as expected failures, like nodes crashing and message loss.

\textbf{(2) Document and communicate the behavior of the system.}
A formal specification can act as compact and unambiguous documentation of system behavior for potential users. This gives developers and users clarity about the guarantees they can expect and rely on. The existence of such a spec also offers a succinct mechanism to communicate changes to these guarantees.
Due to resource constraints on early TEEs~\cite{Costan16}, the consistency guarantees offered by CCF to clients can be subtle, confusing users and paper reviewers alike. In fact, even when consistency guarantees are seemingly simple, they can still be famously difficult to reason about~\cite{Kleppmann15}. We thus wish to formally define these guarantees, communicate them clearly, and check that these hold, even as failures occur.

\textbf{(3) Increased confidence in the implementation \& design.} 
While many verification efforts focus solely on checking the design via reference specs, we also wish to validate that our concrete implementation corresponds to our specs.
If, in the future, there were to be variants or even multiple implementations of CCF, we would like to verify that they are functionally equivalent in key areas.

\textbf{(4) Integrates with the existing codebase.} We do not wish, however, to rewrite CCF for the purpose of verification.
At the time of writing, the implementation is already 63 kLoC in C++. Our support for Intel SGX~\cite{Costan16} (via OpenEnclave SDK~\cite{OE}) constrains us to C++.
Moreover, since the project started seven years ago, we have invested significantly in adding functionality, improving performance, and supporting new hardware (AMD SEV-SNP~\cite{SEV-SNP}).
We want our verification efforts to \textit{improve} upon that existing investment, rather than impose a fresh implementation.

\textbf{(5) Pragmatic and evolves with the implementation over time.} 
CCF is an ongoing project which is continuously growing and evolving, with an average 16 pull requests merged every week.
Since 1.0, the first release to be deployed to production, there have been four further major versions, with minor versions and patches released every 11 days on average.
Any verification effort must thus be continuous, automatic, and sufficiently lightweight to integrate into the existing CI pipelines and software engineering workloads. 
The approach should also be approachable and pragmatic such that anyone contributing to the project can update the specs and debug discrepancies between the implementation and specs.

\subsection{Approach}

Full formal verification of distributed systems has been successfully applied in various research projects~\cite{Wilcox15,hackettCompilingDistributedSystem2023,Hawblitzel17,Hance20}, for instance, by synthesizing an executable implementation from a formally proven spec. Unfortunately, while formal verification is a powerful tool, it requires a significant upfront investment in time and expertise. We thus chose a different approach.  We already used traditional testing techniques (\emph{casual verification}), but wanted to augment them with an approach that was more rigorous and complete. We thus chose to adopt
\emph{smart casual verification}\footnote{Smart casual refers to a style of dress that is neat and stylish, without the expense and discomfort associated with formal attire such as a business suit. Smart casual dress can easily be made more/less formal, for instance, with the addition/removal of a blazer or tie.~\cite{smart-casual}}, a pragmatic yet systematic approach to verification that combines the rigor of formal methods with the easy-of-use and flexibility of more casual methods. We combine a rigorous \tlaplus specification which we \textit{tie} to our existing production implementation using \textit{trace validation}.

\summary{Why \tlaplus?}

We chose \tlaplus for several reasons.
\tlaplus has been successfully utilized to verify the design of a number of production distributed systems~\cite{Newcombe14,tendermint-tla, Schultz22, Yin20,Brooker20}.
Most notably for us, \tlaplus has been used to describe both consistency guarantees~\cite{Hackett23,cosmosdb-tla,tla-lin}, and distributed consensus protocols, namely Paxos~\cite{Lamport98,paxos-tla} and Raft~\cite{raft-tla}.
The existence of the last of these was a significant factor in our decision, as it allowed us to start from a complete spec of Raft and adapt it to our protocol's specificities, rather than from scratch.
Finally, the maturity of \tlaplus and its tools~\cite{DBLP:journals/corr/abs-1912-10633,yu:tlc,Konnov19,Cousineau2012a}, the extensive examples available~\cite{Lamport_TLA_Examples}, and its active community~\cite{tla-group} assured us of ongoing support and resources.  Our increased investment in \tlaplus, following initial successes, was supported by the availability of a recurring, two-day \tlaplus workshop~\cite{10.1007/978-3-031-27534-0_5} to train our team.

However, achieving our goals is not simply a matter of writing some high-level specs as this would not provide any guarantees about the production code itself.
We bridged this gap by applying trace validation to validate implementation traces against the formal specs.
\summary{Why bother to read this paper?}
In this paper, we describe our experience applying smart casual verification to CCF using \tlaplus, focusing on the challenges we faced and how we overcame them.
First, we present our \tlaplus specs of CCF from the perspective of its nodes (\S\ref{sec:consensus}) and its clients (\S\ref{sec:client}).
Next, we present how we validate traces generated from the CCF implementation against our \tlaplus specs (\S\ref{sec:trace}).
Finally, we reflect on our experiences applying smart casual verification to CCF (\S\ref{sec:results}), documenting the six bugs we prevented, and the lessons learned along the way (\S\ref{sec:lessons}).
Both our specifications and implementation are open source and actively maintained\cite{CCF-TLA}.

%% file: 2-ccf.tex
\section{CCF}\label{sec:ccf}

This section provides an overview of the distributed architecture of CCF, focusing on the components that are relevant to our verification efforts.
Interested readers can find a more comprehensive description of CCF in \cite{Howard23}.

At its core, CCF uses state-machine-replication (SMR)~\cite{Schneider90} and trusted execution to offer the abstraction of an always available application that remains robust to attacks, including from the nodes themselves. CCF assumes that neither node operators nor other applications running on the hardware can be trusted. The host, the OS, the hypervisor, the network and persistent storage are all assumed to be corruptible. 
Intuitively, SMR provides clients with the illusion that there is a single server that will, sequentially, execute individual application requests.
SMR achieves this by replicating application logic on a set of nodes, a fraction of which may fail.
The system maintains consistency across nodes by deciding on a totally ordered transaction log.
Formally, SMR guarantees the following (def. from \cite[Fig. 3]{Ongaro14}):

\begin{property}[State Machine Safety]\label{prop:smr}
	If a node has applied a log entry at a given index to its state machine, no other node will ever apply a different log entry for the same index.
\end{property}

Most SMR systems further ensure that the agreed-upon set of transactions will be applied in a way that guarantees \textit{linearizability}~\cite{Herlihy90} (or \textit{strict serializability}): the resulting execution will be equivalent (equal read and write sets) to an execution in which each transaction was executed in sequence, and in an order that matches the order in which they were issued. Though appealing, strict serializability can be costly to enforce.
SMR systems thus often relax this guarantee to read-only transactions specifically.
They offer only \textit{serializability} and allow read-only transactions to read stale state.
CCF is no different: it offers strict serializability for committed read-write transactions and serializability for committed read-only transactions.

Serializability is a gold standard in system design, but is fairly pessimistic: a client must wait until a transaction has committed to learn any information about that transaction and its effects.
This design was unfortunately at odds with CCF's initial SGX-related design constraint, which precluded keeping potentially large amounts of application-defined responses in the limited amount of in-enclave memory (128 MB).
CCF thus provides configuration settings that clients can use to achieve good performance.
They are useful in practice, but make it more challenging to formalize the consistency guarantees that users can expect.

In CCF, the leader node executes transactions as soon as they are received, and prior to them being replicated to other nodes.
The leader then directly replies to the client with the result of the transaction without waiting for confirmation that the request has been replicated.
As a consequence, a leader failure can cause the transaction to fail, even after a response has been returned to the client.
Clients can then, on a per-response basis, decide whether they wish to wait for the transaction to be committed before proceeding or not.
Either way, this cuts down the number of open client connections and pending responses, reducing memory footprint significantly. 

Transactions in CCF can be in one of the following client-observable states:
\textsc{Committed}, \textsc{Pending}, or \textsc{Invalid}.
A transaction is \textsc{Committed} if it has been replicated by the leader to a majority of nodes in its current term.
A transaction is \textsc{Pending} if it has been executed but not yet replicated. 
If the leader fails before replication is complete, CCF will mark the transaction as \textsc{Invalid}.

\textsc{Pending} transactions, can eventually become \textsc{Invalid}, but cannot return arbitrary results.
They provide a guarantee akin to \textit{fork-linearizability}~\cite{mazieresBuildingSecureFile2002} (or \textit{fork sequential consistency}~\cite{Brandenburger17} when considering read-only transactions).
A pending transaction observes a prefix of committed transactions and a sequence of pending transactions.
Leader failures may cause the system to fork and generate multiple (locally linearizable) sequences of pending transactions.
Only one forked sequence will eventually commit, thus ensuring that the set of committed transactions remains linearizable.
All other sequences will be marked as invalid. In other words, if a pending transaction commits, the result it returned to clients is guaranteed to have been linearizable.

CCF makes extensive use of timestamps to help clients understand when transactions transition from \textsc{Pending} to \textsc{Committed} or \textsc{Invalid}.
Each transaction is associated with a unique transaction identifier, consisting of a lexicographically ordered pair $\langle t.i \rangle$ of term $t$ and log index $i$. %
The client can use this ID to quickly understand the system state.
CCF, for instance, enforces \textit{timestamp ordering}: if $txid < txid'$ and the two transactions are committed, then the transaction with $txid$ executed before $txid'$.
Clients can further use this ID to learn the state of not only this transaction, but its ancestors. For instance, CCF guarantees that:  
\begin{property}[Ancestor Commit]\label{prop:committed}
	If $\langle t.i \rangle$ is committed then any transaction $\langle t.j \rangle$ where $j \leq i$ is also committed.
\end{property}

Ensuring that CCF does indeed provide these specific guarantees requires care and adapting existing linearizability specs, which only reason about committed operations.
They support neither reasoning about forks nor timestamp properties.

\summary{background on consensus}
To tolerate node crashes and network asynchrony, SMR requires a crash fault-tolerant distributed consensus protocol or equivalent~\cite{Chandra96}, such as Multi-Paxos~\cite{Lamport98} or Raft~\cite{Ongaro14}, to agree on a total-ordered transaction log.
While the terminology varies, such protocols are typically leader-based and operate by electing one of the nodes to be the \emph{leader} while the other nodes are \emph{followers}.
The leader is responsible for proposing new transactions for the log and the followers are responsible for replicating them.
The leader will only consider a transaction to be committed once a strict majority of nodes have replicated the transaction at the same position in the log.
When the leader fails, a new leader is elected from the remaining nodes and the protocol continues.
We use \emph{terms} to distinguish between the different periods of leadership and there should be at most one leader per term.

\summary{background on OG raft} It is the responsibility of the leader-based consensus protocol to ensure that the new leader has knowledge of all previously committed transactions.
In Raft, this is achieved by requiring followers to become \emph{candidates} (transition \circlelabel{1} in~\autoref{fig:state-transitions}) before they can become leaders.
A candidate will only become a leader (transition \circlelabel{2} in~\autoref{fig:state-transitions}) if it can obtain a strict majority of votes from the other nodes.
A node will only vote for a candidate if it has not already voted in the candidate's term and the candidate's log is at least as up-to-date as its own log.
The former ensures that at most one leader can be elected per term and the latter ensures that the new leader has knowledge of all committed transactions from previous terms.
A more comprehensive description of Raft can be found elsewhere~\cite{Ongaro14}.

\subsection{What makes CCF's distributed consensus protocol interesting?}\label{subsec:ccf/consensus}

\summary{Why is CCF's consensus interesting enough to verify?}
CCF uses a custom consensus protocol, which evolved from Raft. 
Now we will outline some of the ways that the consensus protocol in CCF today differs from the description of the Raft protocol given the original paper.
While some of these modifications may seem small, the combined effect is significant and the interactions between them have proven complex and subtle, leading to the bugs we later describe (\S\ref{sec:results}).

\textbf{Signature transactions}
Offline log integrity and transaction provenance are key requirements for CCF, neither of which is provided by Raft, nor by the AEAD mechanism used inside the CCF network.
The offline guarantees crucially enable external audit, and disaster recovery.
To implement them efficiently, CCF utilizes \emph{signature transactions}, which include the root of a Merkle tree~\cite{Merkle87} over the whole log thus far, signed by the current leader. 
A transaction in the log is not considered committed unless a subsequent signature has been committed.

\textbf{Messaging not RPCs}
CCF does not use RPCs to communicate between nodes, and instead it uses a uni-directional messaging layer.
When a node receives a response to a message it has sent, it does not know which message the response corresponds to as we do not assume reliable or in-order delivery.
Raft uses two main message types: \textsc{AppendEntries} (AE) and \textsc{RequestVote} (RV).
AE messages are used to replicate transactions and RV messages are used to elect a leader.
In the case of RVs, the term in the response is sufficient to handle the reply.
In the case of AEs, the response in CCF contains an additional field, \textsc{LastIndex}.
For positive responses to AE messages (AE-ACK), this is the index of the last transaction in the follower's log.

\textbf{Optimistic acknowledgement}
In Raft, the leader replicates transactions to its followers using AE requests. 
The leader maintains a \textsc{NextIndex} for each node to record which log entry to send next to the follower. 
This index is updated when a follower responds positively to an AE request (AE-ACK).
If a leader receives a new transaction from a client, before it has received a response from a follower, it must either (1) wait for the reply for the previous transaction, potentially impacting liveness and performance as AEs cannot be pipelined, or (2) send both transactions in the next AE request, even though the follower is likely to already have the first transaction. 
CCF avoids this problem by allowing the leader to update the index, known in CCF as the \textsc{SentIndex}, as soon the AE message is sent. This therefore means that if the leader receives a negative response to its AE request (AE-NACK), it might need to rollback the \textsc{SentIndex}.

\textbf{Express node catch up}
AE messages include a previous log index and term, which allows a follower to determine if it diverged from the leader.
If a follower does not have the previous log index and term, either because it does not have, or has a different transaction at that index, it then responds with an AE-NACK.
CCF uses an express catch up mechanism, where a leader makes a conservative estimate of how far behind a follower is, and sends a batch of transactions to the follower.

\begin{figure}
	\centering
	\includegraphics[scale=0.65]{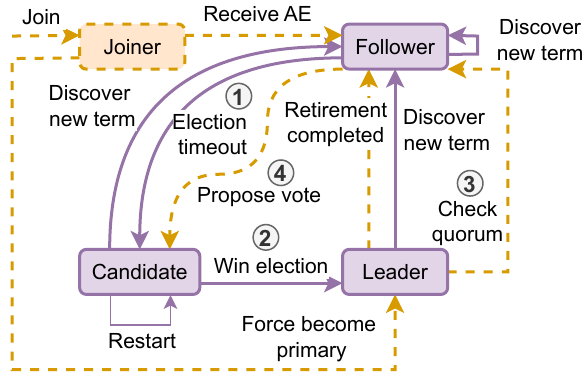}
	\caption{State transitions for CCF's consensus protocol. Solid boxes and lines show Raft's original states and transitions, CCF's additional states and transitions are shown with dashed lines.}\label{fig:state-transitions}	
\end{figure}

\textbf{Partition leader step down}
A known limitation of the base Raft protocol is that partial/asymmetric network partitions can cause a loss of liveness~\cite{Jensen21,raft-liveness}.
For instance, if a leader can no longer make progress because it cannot receive messages from the other nodes, it continues to send AE heartbeats to followers, preventing them from timing out and from electing a new leader who can make progress.
CCF implements a known extension~\cite[pg. 69]{diego} to the Raft protocol, referred to as \emph{CheckQuorum}, where a leader steps down if it does not hear back from a quorum of nodes within a specified time period (transition \circlelabel{3} in~\autoref{fig:state-transitions}).

\textbf{Bootstrapping to retirement}\label{subsec:ccf/reconfiguration}
Reconfiguration is the process by which the set of nodes participating in consensus can be changed.
Raft support two reconfiguration protocols: \emph{joint consensus}, described in the original paper~\cite{Ongaro14}, and single-server reconfiguration, subsequently described in \cite{diego}.
CCF's reconfiguration protocol is more similar to the former. 
Reconfigurations are recorded in CCF's log with \emph{configuration transactions}, and are therefore ordered in the same total order.
Logs always begin with an initial singleton configuration transaction followed by a signature transaction.
To change the configuration, the leader proposes a new configuration transaction specifying the new set of nodes, which may be different in cardinality to the current set of nodes and may or may not be disjoint.
To commit this transaction, the leader must obtain a quorum of AE-ACKs from both the previous and new configurations.
Once the transaction is committed, the leader no longer needs quorum agreement from the previous configuration.
If a node has been removed from the configuration, we refer to it as \emph{retiring}.
In order to complete its retirement and permanently switch off the node, a \emph{retirement transaction} must be committed to ensure that any future leader will know that the reconfiguration which removed the node has been committed, and thus the node will never be needed.
CCF also adds a message to the protocol, \emph{ProposeVote} which is utilized by a retiring leader to nominate a successor, fast-tracking the usual leader election process (transition \circlelabel{4} in~\autoref{fig:state-transitions}).

%% file: 3-tla.tex
\section{Primer on \tlaplus{}}\label{sec:tla}

\summary{\tlaplus{} the language}
\tlaplus~\cite{SpecifyingSystems,Lamport94} is a formal modeling language widely used to verify concurrent and distributed systems.
It is easy to learn and use, as well as agnostic about system frameworks or implementation languages.

\tlaplus{} is a variant of linear temporal time logic with only two operators, \emph{Always} ($\Box$) and \emph{Eventually} ($\diamond$).
A system is defined by a set of \emph{behaviors}, each a sequence of pairs of states called \emph{actions}, beginning at one of the system's initial states.
A \emph{state} is an assignment of values to variables.  Syntactically, \tlaplus{} describes a system's state machine using a canonical (temporal) formula $Init \wedge \Box[Next]_{vars} \wedge L$.
Here, $Init$ is a predicate defining the system's set of initial states.
The system's next-state relation $Next$ is a first-order logic formula that is usually decomposed into a disjunct of actions which relates the values of the variables in the current state to the ones in the successor state.
\textsc{CheckQuorum} (Listing~\ref{lst:consensus}), for instance, states that a node $i$ can abdicate as a leader and become a follower; we change the value of the variable \textit{role} to $Follower$ in the successor state, while the values of the other variables remain unchanged.\footnote{The value of the variable \textit{role} is a mapping from all node identifiers to their roles such as \emph{leader}, \emph{follower}, and \emph{candidate}.}
The tuple, $vars$, represents the spec's variables, and $[Next]_{vars}$ stipulates that either $Next$ is true, or the variable in $vars$ do not change.
This asserts that \tlaplus{} specs are stuttering-insensitive, allowing a spec to always be refined by a more detailed, low-level one.
The optional formula $L$ is used to assert fairness, i.e., constraints on the system's behavior that ensure that certain actions eventually occur.
Composition of actions allows us to change the \emph{grain of atomicity} by defining more coarse-grained behaviors~\cite[\S7.3]{SpecifyingSystems}.
Concretely, the composition $A \cdot B$ states that the two actions happen atomically; the intermediate state between $A$ and $B$ is not observable.

\summary{TLC}

We also state desired safety (\emph{something bad never happens}) and liveness (\emph{something good eventually happens}) properties in \tlaplus{}. For example, the safety properties \textsc{LogInv}, \textsc{AppendOnlyProp}, and \textsc{MonoLogInv} (Listing~\ref{lst:consensus}, described in \S\ref{sec:consensus}) assert properties of the \textit{log}.
Properties are checked and verified using, random state space exploration (simulation), model checking, or theorem proving.
TLC~\cite{yu:tlc}, an explicit-state model checker, verifies that a \emph{finite} model of a spec satisfies its properties by enumerating all reachable states.
A symbolic model-checker~\cite{Konnov19} is especially well-suited for the verification of inductive invariants of finite systems.
The \tlaplus{} Proof System~\cite{tlapm, Cousineau2012a} mechanically verifies a deductive proof of the properties of an infinite system.
These tools complement each other, enabling a combination of model-checking and deductive proofs to verify specs.
This approach allows for varying levels of verification depth, from push-button model checking to fully mechanized safety and liveness proofs~\cite{konnovSpecificationVerificationTLA2022}.
Our companion paper~\cite[\S2.1]{cirsteaValidatingTracesDistributed2024} provides a more detailed summary of \tlaplus{}.

%% file: 4-consensus.tex
\section{Distributed Consensus Specification}\label{sec:consensus}

\begin{figure}
	\centering
	\includegraphics[scale=0.5]{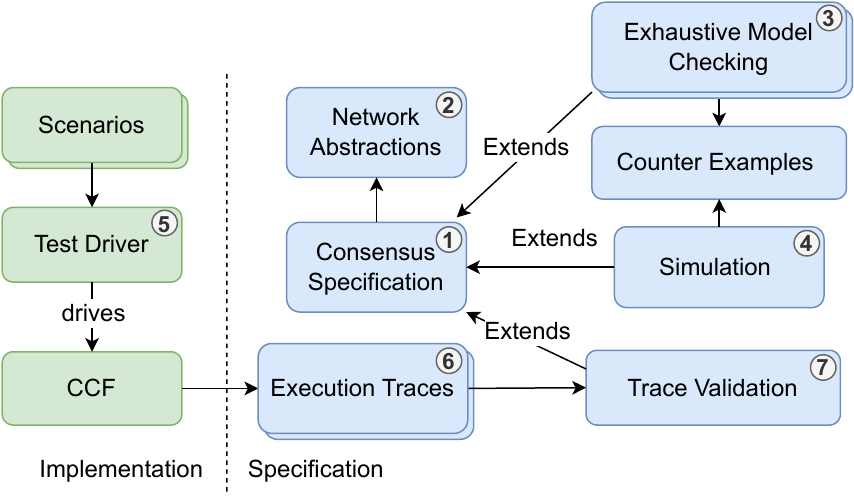}
	\caption{Components of our verification architecture for consensus}\label{fig:compo}
\end{figure}

\renewcommand{\figurename}{Listing}
\begin{figure}
	\raggedright
	\input{code/consensus.tex}
	\caption{Excerpt of our consensus spec. 
	}\label{lst:consensus}
\end{figure}

\autoref{fig:compo} summarizes our consensus verification architecture.
The consensus specification, shown as \circlelabel{1} in \autoref{fig:compo}, consists of 17 actions to describe the transitions over 13 variables.
The first 12 variables are local and track consensus state (\textsc{currentTerm}, \textsc{log}, \textsc{commitIndex}, \dots).
The last variable instead tracks the set of in-transit messages, allowing support for different network abstractions (ordered/unordered delivery, etc.) and is shown as \circlelabel{2} in \autoref{fig:compo}.
Each action models a node taking a single step within the protocol, updating its local state and optionally the collection of in-transit messages, (i.e. \textsc{CheckQuorum} in Listing~\ref{lst:consensus}). 
Actions can either be initiated from a message receipt or by a node itself.
Actions in the latter class include adding a signature transaction, stepping up as a candidate or stepping down as a leader.
Such actions are always enabled, to model the fact that we make no assumptions about clock synchrony; each node's opinion of the progression of time is independent.
Later we discuss how action-weighted simulation enabled us to find bugs in the prototype despite the state space explosion inherent with such an approach.
The spec is parameterized by the set of nodes available in the service.
The initial states of the spec include every non-empty subset of nodes in the initial configuration with any node in that initial configuration as an initial leader.

Our key correctness property is State Machine Safety (Property~\ref{prop:smr}) which we check with invariant \textsc{LogInv} and action property \textsc{AppendOnlyProp} (Listing~\ref{lst:consensus}).
\textsc{LogInv} states that all pairs of committed logs must be consistent and \textsc{AppendOnlyProp} states that each node can only extend its committed log.
\textsc{LogInv} checks for safety violations across nodes (in space) and \textsc{AppendOnlyProp} checks for safety violations within a node (in time).
We also checked a further 27 invariants/properties to ensure the correctness of our consensus protocol and of our understanding (see \cite{CCF-TLA}).
One such example is \textsc{MonoLogInv} (Listing~\ref{lst:consensus}), which is a stronger variant of the property that log terms are monotonically increasing which we depend upon extensively.
More concretely, \textsc{MonoLogInv} states that terms in the log can only increase after a signature and remains the same otherwise.

Our consensus spec is unbounded, there is always the possibility of a new transaction being proposed or a timeout triggering a new election.
To exhaustively model check our spec, we extend our consensus spec, shown as \circlelabel{3} in \autoref{fig:compo}, to restrict the state space by adding additional constraints to the actions limiting the max term, number of client requests, and the sequence of reconfigurations.
We found exhaustive model checking too time-consuming to run with all but the strongest state constraints in our CI pipeline, particularly after the spec was updated to describe reconfiguration accurately (discussed further in \S\ref{sec:lessons}).
We therefore developed an extension to our spec for simulation as a lightweight alternative to exhaustive state exploration.
Our simulation spec, shown as \circlelabel{4} in \autoref{fig:compo}, takes a time quota and explores as many behaviors as possible, up to a given depth, within that time.
To expand the coverage of simulation, specifically to explore behaviors where the system exhibits more forward progress, we manually weighted failure actions to reduce the likelihood of them being chosen.
We also implemented Q-Learning proposed by~\cite{mukherjeeLearningbasedControlledConcurrency2020} in TLC to automatically weight actions to increase coverage of simulation.
However, we were unable to find the right set of variables as input to Q-Learning's state hash function $\mathcal{H}$ that achieved better coverage than manual weighting.

%% file: code/consensus.tex
\begin{tla}
CheckQuorum(i) ==
    /\ role[i] = Leader
    /\ role' = [role EXCEPT ![i] = Follower]
    /\ UNCHANGED <<currentTerm, log, votedFor, ...>>

LogInv == \A i, j \in Nodes :
    \/ IsPrefix(Committed(i),Committed(j)) 
    \/ IsPrefix(Committed(j),Committed(i))
        
AppendOnlyProp ==
    [][\A i \in Nodes :
        IsPrefix(Committed(i), Committed(i)')]_vars

MonoLogInv == \A i \in Nodes : log[i] # <<>> => 
    \A k \in 1..Len(log[i])-1 :
        \/ log[i][k].term = log[i][k+1].term
        \/  /\ log[i][k].term < log[i][k+1].term
              /\ log[i][k].contentType = Signature
\end{tla}
\begin{tlatex}
\@x{ CheckQuorum ( i ) \.{\defeq}}%
\@x{\@s{16.4} \.{\land} role [ i ] \.{=} Leader}%
 \@x{\@s{16.4} \.{\land} role \.{'} \.{=} [ role {\EXCEPT} {\bang} [ i ] \.{=}
 Follower ]}%
 \@x{\@s{16.4} \.{\land} {\UNCHANGED} {\langle} currentTerm ,\, log ,\,
 votedFor ,\, \.{\dots} {\rangle}}%
\@pvspace{8.0pt}%
\@x{ LogInv \.{\defeq} \A\, i ,\, j \.{\in} Nodes \.{:}}%
\@x{\@s{16.4} \.{\lor} IsPrefix ( Committed ( i ) ,\, Committed ( j ) )}%
\@x{\@s{16.4} \.{\lor} IsPrefix ( Committed ( j ) ,\, Committed ( i ) )}%
\@pvspace{8.0pt}%
\@x{ AppendOnlyProp \.{\defeq}}%
\@x{\@s{16.4} {\Box} [ \A\, i \.{\in} Nodes \.{:}}%
 \@x{\@s{20.5} IsPrefix ( Committed ( i ) ,\, Committed ( i ) \.{'} ) ]_{
 vars}}%
\@pvspace{8.0pt}%
 \@x{ MonoLogInv \.{\defeq} \A\, i \.{\in} Nodes \.{:} log [ i ] \.{\neq}
 {\langle} {\rangle} \.{\implies}}%
\@x{\@s{16.4} \A\, k \.{\in} 1 \.{\dotdot} Len ( log [ i ] ) \.{-} 1 \.{:}}%
 \@x{\@s{20.5} \.{\lor} log [ i ] [ k ] . term \.{=} log [ i ] [ k \.{+} 1 ] .
 term}%
 \@x{\@s{20.5} \.{\lor}\@s{4.1} \.{\land} log [ i ] [ k ] . term \.{<} log [ i
 ] [ k \.{+} 1 ] . term}%
\@x{\@s{32.8} \.{\land} log [ i ] [ k ] . contentType \.{=} Signature}%
\end{tlatex}

%% file: 5-consistency.tex
\section{Client Consistency Specification}\label{sec:client}

Following our success with the consensus spec and a discussion regarding the linearizability of read-only transactions~\cite{ccf-5636}, we decided to formalize the possible externally visible behaviors of a CCF service in \tlaplus{} to better understand the guarantees provided.
Our aim with this spec was to keep it as high-level as possible, focusing solely on the possible safe interactions between clients and the service. 
By design this spec does not model the internal details of the service itself such as the state of individual nodes or the messages exchanged between nodes as such low level details are already modeled by our consensus spec.

Our consistency spec uses just two variables.
The first is \textsc{History}, an append-only sequence which records the messages exchanges between clients and the services.
Five messages are supported, read-only/read-write transaction requests and responses as well as a transaction status messages.
Note that since we are focused on safety, we omit messages that cannot impact correctness but can increase the state space, for instance, we do not track when a client requested a transaction status, nor do we track status responses of \textsc{Pending}.
In order to stress the consistency guarantees, we modelled an application where all transactions operate on a simple value, reading the current value, appending a new identifier to the value and writing back the new value.
All transactions conflict and each transaction observes every transaction that has been executed before it.
To succinctly record the current state of the service we use \textsc{LogBranches}, an append-only two-dimensional sequence, where the sequence at index $i$ corresponds with the local log of the leader of term $i$ if such a node still exists.
This representation does not therefore need to be parameterized over the number of nodes in the service and usefully models the fact that there can be multiple leaders at one time (although they will have different terms).

\renewcommand{\figurename}{Listing}
\begin{figure}
	\raggedright
    \input{code/historyinvars.tex}
	\caption{Two properties over histories checked by our consistency spec. 
    }\label{lst:historyinv}
\end{figure}

To populate these histories, we defined a set of actions to describe how the history can be extended depending on the log branches.
Initially, the history and log branches are empty.
The possible actions are: append a transaction request or response to the history, execute a transaction by appending it to any log branches, append a transaction status to the history, or starting a new log branch to simulate leader election.
Note that when a transaction is executed, it can be appended to any log branch, this simulates the fact that a transaction can be handled by any node that believes itself to be the leader, even if it is not the latest leader.
At any time, a new log branch can be started.
This new log branch can be any prefix of any existing log branches, provided it includes the last committed transaction.
Having established an approach to generate histories, we formalized in \tlaplus{} the properties that we expect to hold over these histories and check them with TLC.
For instance, \textsc{PrevCommittedInv} (Listings~\ref{lst:historyinv}) formalizes Property~\ref{prop:committed} by stating that for any pair of transaction status responses from the same term, if the one with the greater (or equal) index is \textsc{Committed}, then the other status response must also be \textsc{Committed} (as \textsc{Pending} states are not modelled).
\textsc{ObservedRoInv} (Listings~\ref{lst:historyinv}) states that if a committed read-write transaction received an initial response (event $i$), before a committed read-only transaction was started (event $j$) then the read-only transaction response (event $k$) must observe the read-write transaction.
Note that \textsc{RwResCommitIndexes} is the set of indexes for all read-write transaction response events, filtered to include only transactions that were subsequently committed.
\textsc{RoReqCommitIndexes} and \textsc{RoResCommitIndexes} are similarly defined for read-only transaction requests and responses.

%% file: code/historyinvars.tex
\begin{tla}
PrevCommittedInv ==
  \A i, j \in {x \in DOMAIN history: history[x].type = Status}:
      /\ history[i].status = Committed
      /\ history[i].term = history[j].term
      /\ history[j].index <= history[i].index
      => history[j].status = Committed

ObservedRoInv ==
  \A i \in RwResCommitIndexes :
      \A j \in RoReqCommitIndexes :
          \A k \in RoResCommitIndexes :
              history[k].tx = history[j].tx /\ i < j
              => Contains(history[k].observed, history[i].tx)
\end{tla}
\begin{tlatex}
\@x{ PrevCommittedInv \.{\defeq}}%
 \@x{\@s{8.2} \A\, i ,\, j \.{\in} \{ x \.{\in} {\DOMAIN} history \.{:}
 history [ x ] . type \.{=} Status \} \.{:}}%
\@x{\@s{8.2} \.{\land} history [ i ] . status \.{=} Committed}%
\@x{\@s{8.2} \.{\land} history [ i ] . term \.{=} history [ j ] . term}%
\@x{\@s{8.2} \.{\land} history [ j ] . index \.{\leq} history [ i ] . index}%
\@x{\@s{8.2} \.{\implies} history [ j ] . status \.{=} Committed}%
\@pvspace{8.0pt}%
\@x{ ObservedRoInv \.{\defeq}}%
\@x{\@s{8.2} \A\, i \.{\in} RwResCommitIndexes \.{:}}%
\@x{\@s{12.29} \A\, j \.{\in} RoReqCommitIndexes \.{:}}%
\@x{\@s{16.4} \A\, k \.{\in} RoResCommitIndexes \.{:}}%
 \@x{\@s{20.5} history [ k ] . tx \.{=} history [ j ] . tx \.{\land} i \.{<}
 j}%
 \@x{\@s{20.5} \.{\implies} Contains ( history [ k ] . observed ,\, history [
 i ] . tx )}%
\end{tlatex}

%% file: 6-trace.tex
\section{Trace Validation}\label{sec:trace}

Conceptually, trace validation checks that every observed implementation trace matches a behavior of the system's high-level specification.
We must therefore first collect implementation traces before validating them against our high-level consensus \tlaplus specification. 
A more formal discussion of trace validation, including insights from applying trace validation to systems other than CCF, is provided in~\cite{cirsteaValidatingTracesDistributed2024}.

\subsection{Trace Collection}\label{subsec:trace/existing}

When we began the trace validation work, CCF already had extensive unit, functional, and end-to-end testing.
Unit testing of CCF's consensus layer consisted of approximately 1.6 kLoC.
End-to-end testing covered complex scenarios such as reconfigurations, node failover, and network partitions, spread over more than 2 kLoC of Python tests and infrastructure code.
Consensus functional testing was done through a scenario driver, shown as \circlelabel{5} in \autoref{fig:compo}, that serialized execution deterministically across nodes, and isolated the consensus layer by mocking unrelated CCF components, such as governance, networking etc.
This driver allowed the injection of network faults such as partitions, delays, reorderings, and message loss, and provides observability.
Core correctness invariants and properties were checked at designated execution steps in 13 manually written scenario tests exercising replication, election, and reconfiguration under controlled fault conditions.
Additionally, an initial prototype to fuzz-test the consensus layer through generated inputs and faults was developed but ultimately abandoned since it failed to generate interesting behaviors that would achieve satisfactory coverage.

\label{subsec:trace/whenlog}
The driver allows us to replace the node's wall clocks with a single global clock.
If this global clock had not been possible, a distributed clock, such as a Lamport or vector clock, would have also provided the necessary event ordering to establish the happen-before relationships.
However, a distributed clock would have required changes to the network message format.
We enhanced system observability by incorporating an additional 15 log statements to capture consistent system states at well-defined, side-effect-free linearization points including (i) the sending and receipt of network messages, and (ii) transitions in a node's high-level state, such as moving from candidate to leader\footnote{State changes are logged immediately after acquiring a global lock.}.
It is important to note that the driver logs only those values that remain constant in space. For instance, the driver records the length of the logs but not the log entries themselves, which would become impractically large.
Still, the logging is disabled at compile time for production builds, and thus does not impact CCF's performance.

Before validation, implementation traces, shown as \circlelabel{6} in \autoref{fig:compo}, are preprocessed to exclude and de-duplicate events from the initial bootstrapping phase of a CCF network, as this phase is not modeled in our consensus spec.

\subsection{Consensus Trace Validation}\label{subsec:trace/idea}

Trace validation, shown as \circlelabel{7} in \autoref{fig:compo}, ensures that an implementation trace is consistent with the specification. More formally, it verifies whether the set of behaviors $\mathcal{T}$, which encapsulates the values and events from a trace, intersects with the set of behaviors $\mathcal{S}$ derived from the high-level spec, thereby checking that $\mathcal{T} \cap \mathcal{S} \neq \emptyset$.
While TLC can construct $\mathcal{S}$, it cannot directly generate $\mathcal{T}$ from the trace.
Thus, we write a new \tlaplus spec, \textit{Trace}, reusing many actions of the definitions from the high-level spec.  However, the actions are only enabled iff the current event in the trace matches an action. Likewise, the actions are parameterized by the values taken from the trace, effectively constraining successor states.

\renewcommand{\figurename}{Listing}
\begin{figure}
	\raggedright
	\input{code/traceccfraft.tex}
	\caption{Excerpt of the \textit{Trace} spec for trace validation.}
	\label{lst:traceccfraft}
\end{figure}

Consider for example \textit{IsSendAppendEntries} (Listing \ref{lst:traceccfraft}). This action is enabled iff the current line of the trace, denoted by \textit{ln}, is a \textit{sndAE} event, and the \textit{commitIndex} of the sending node (denoted by \textit{ln.snd}) matches the trace's \textit{commit\_idx}.
The action then reuses the definition of the \textit{AppendEntries} action from the consensus spec, parameterized by \textit{ln.snd} and \textit{ln.rcv} from the trace.
Given that the number of entries in the high-level action \textit{AppendEntries} is chosen nondeterministically within defined limits, \textit{IsSendAppendEntries} determines the successor state by asserting the existence of an additional \textit{AppendEntriesRequest} in the network with a matching number of entries.
Note that while the consensus spec modeled the network as a set of messages, thereby leaving the variable upon a resend of an \textit{AppendEntriesRequest} unchanged, \textit{Trace} redefines the value of the variable to be a multi-set.
This allowed the spec to account for the addition of messages in the network, even during resend events.
Subsequently, this approach to address the impedance mismatch was expanded to verify, with TLC, the impact of various message delivery guarantees, such as ordering, duplication, and other message loss patterns.

\subsubsection{Aligning Grains Of Atomicity}\label{subsec:trace/atomicity}

The granularity of some actions in the consensus spec did not align with the granularity of events in the traces.
The action \textit{IsRcvAppendEntries} outlines the alignment of such different \emph{grains of atomicity}.
Like many Raft implementations, CCF minimize network round trips by piggybacking \emph{term} updates on \textit{AppendEntries} messages.
This optimization was not reflected in the consensus spec. 
Instead, the spec modeled term updates with an action that increases the term upon a pending \textit{AppendEntriesRequest}, while leaving the variable unchanged.
In effect, a term update might nondeterministically occur before the receipt of the \textit{AppendEntriesRequest}.
To reconcile these different grains of atomicity, we composed the actions \textit{UpdateTerm} and \textit{HandleAppendEntriesReq}, allowing them to occur atomically, i.e., in a single action.

Another important application of action composition addresses events that are omitted from the trace, such as losing messages.
Although our consensus specs explicitly modeled losing messages, message loss was not recorded in the trace.
Therefore, to account for faults at any step of a behavior, we composed an \textit{IsFault} action with \textit{Trace}'s next-state relation.
Conversely, aligning a single high-level action with multiple implementation events is addressed by introducing finite stuttering that does not change the high-level variables.
For instance, the action \textit{IsSendAppendEntriesResponse} is enabled iff \textit{ln} is a \textit{sndAER} event.  However, it leaves the high-level variables unchanged.

\subsubsection{Deriving Consensus Trace Validation}\label{subsec:trace/workflow}

We began to derive \textit{Trace} by mapping the trace of a straightforward happy-path test line by line to the consensus spec.
To proactively catch discrepancies, we added as many assertions as possible to \textit{Trace}.
Whenever we discovered discrepancies, we investigated them by examining relevant sections of the implementation's source code.
Adding cross-references between the implementation and the spec proved useful, particularly when the terminology—--such as variable and function names—--differed between the two.
We further debugged the \textit{Trace} using the \tlaplus debugger~\cite{kuppeTLADebugger2023} in tandem with implementation debugging.
Upon detecting discrepancies, we corrected either the implementation or the spec, subsequently rerunning verification on the revised consensus spec, and executing tests on the updated implementation.

\subsection{Debugging Discrepancies}\label{subsec:trace/debugging}

Bogus logging, incorrect mappings from implementation to spec state, or true discrepancies between the spec and the implementation, resulted in $\mathcal{T} \cap \mathcal{S} = \emptyset$, i.e., the verdict that a trace is invalid.
In either case, contrary to ordinary model checking, a failure to validate a trace had no counterexample.
However, the behaviors within~$\mathcal{T}$ helped explain why a trace failed to validate.
We typically compared the final state of the longest behaviors and the corresponding line in the trace to identify the source of the mismatch.
The \tlaplus debugger was instrumental in this process, as it allowed us to step through the evaluation of formulas and compare variables at the current and successor states with the trace values.
To determine if \textit{Trace} is overly restrictive, we implemented a new \textit{unsatisfied breakpoint}. It triggers for each state in $\mathcal{T}$ that is found to be unreachable.
Furthermore, $\mathcal{T}$ can be visualized as a graph that not only includes all unreachable states but also references the subformula responsible for each state being unreachable.

\subsection{Scalability of Trace Validation}\label{subsec:trace/scalability}

The cardinality of $\mathcal{T}$, the set of potential system behaviors, can become prohibitively large due to nondeterminism resulting from incomplete traces.
Recognizing that it suffices to find a single behavior in the intersection of $\mathcal{T}$ and $\mathcal{S}$ to check the validity of a trace, we implemented depth-first search (DFS) in TLC.
This method mitigated the issue of state-space explosion, making trace validation orders of magnitude faster compared to enumerating all behaviors with breadth-first search (BFS).
For instance, validating a trace against our consistency spec started to take less than a second using DFS, compared to about an hour with BFS.%

\subsection{Trace Validation Effort}\label{subsec:trace/effort}

The enhancements to the test driver and the addition of detailed logging were completed in approximately one day.
The effort to derive a version of the \textit{Trace} spec that validated the majority of the traces required approximately two engineer-months, spread over four months.
The primary tasks included enhancing the TLC model checker to support trace validation, which involved implementing support for action composition, DFS, improved debugging support, and visualizing the state graph.
The second major tasks was diagnosing if the root cause of discrepancies arose from bugs in the reverse-engineered spec, the implementation, or both.
This frequently required consulting the original Raft paper and discussions with the CCF experts to elucidate differences between Raft and CCF.
In this context, the shared vocabulary developed during the \tlaplus workshops, along with \tlaplus counterexamples from trace validation, simulation, and model checking proved invaluable.
The third major task involved finding modeling patterns to bridge impedance mismatches between the high-level \tlaplus design and low-level implementation.
Writing the 400 LoC \textit{Trace} spec itself was a minor task.
The introduction of trace validation resulted in 88 fine-grained commits to the \textit{Trace} spec, while the consensus spec underwent 107 changes.

The discovery of a serious safety bug through trace validation (\emph{Commit advance on AE-NACK} \S\ref{results/bugs/commit_advance_on_aenack}) led to increased investment in trace validation.
Substantial changes were made to the consensus spec to accurately reflect the implementation. 
For example, the bootstrapping of a CCF network was modeled with greater fidelity.
Additionally, the spec was expanded to include all node states, especially those related to node retirement.
These comprehensive changes necessitated substantial revisions to the test driver and the development of new tests. This uncovered a serious liveness bug (\emph{Premature node retirement} \S\ref{results/bugs/premature_node_retirement}).
Once the consensus spec was validated to accurately mirror the implementation, we transitioned to a spec-driven development, wherein the spec served as the source of truth.
Notably, this phase included the integration of the \emph{ProposeVote} messages (described in \S\ref{sec:consensus}; used for transition \circlelabel{4} in~\autoref{fig:state-transitions}).

Later, trace validation was also applied to the consistency spec, requiring significantly less effort.
The consistency spec was considerably less complex, written with the implementation in mind, and we had already gained experience validating our consensus spec.
Moreover, TLC had already been enhanced to support trace validation.
No instrumentation of the CCF source code was required for consistency trace validation. Instead, the implementation state was observed by making calls to the system's REST API.
As with consensus specification, we had to address impedance mismatches.
For instance, the consistency spec assumed knowledge of the transactions of all clients, whereas a trace is limited to the transactions of a single client.
This required defining a TLA+ action in the specification to reconstruct all transactions based on observed transaction IDs.
Yet, applying trace validation to consistency was almost entirely carried out by the CCF experts, with minimal involvement from the formal methods expert.
The effort required to apply trace validation to the consistency spec was approximately one engineer-week, spread over a two-week period.\footnote{The work on trace validation has been tracked in milestones 18 and 20 at \url{https://github.com/microsoft/CCF/milestones/.}}

%% file: code/traceccfraft.tex
\begin{tla}
VARIABLE ln

IsSendAppendEntries ==
  \* Enablement conditions on current state.
  /\ IsEvent(ln, "sndAE")
  /\ commitIndex[ln.snd] = ln.commit_idx
  /\ ...
  \* High-level spec actions.
  /\ AppendEntries(ln.snd, ln.rcv)
  \* Assertions on successor states.
  /\ \E m \in Network!Messages':
      /\ Network!OneMoreMessage(m)
      /\ IsAppendEntriesRequest(m, ln)
  /\ ...

IsRcvAppendEntries ==
  \* Enablement conditions on current state.
  /\ IsEvent(ln, "recvAE")  /\  ...
  \* High-level spec actions.
  /\ \E m \in Network!MessagesToFrom(ln.rcv, ln.snd):
      IsAppendEntriesRequest(m, ln) /\
      \/ HandleAppendEntriesReq(ln.rcv, ln.snd, m)
      \* Impl optimization: Piggyback term on AppendEntries.
      \/ UpdateTerm(ln.rcv, ln.snd, m) \cdot 
              HandleAppendEntriesReq(ln.rcv, ln.snd, m)
      \/ ...     
  \* Assertions on successor states.
  /\ ...

IsSendAppendEntriesResponse ==
  \* Enablement conditions on current and successor state.
  /\ IsEvent(ln, "sndAER")  /\  ...
  \* High-level spec actions.
  /\ UNCHANGED vars

IsFault == \E s \in MultiPowerset(network): 
    network' = s /\ UNCHANGED AllVarsExceptNetwork

Spec == Init /\ [][IsFault \cdot Next]_<<vars, ln>>
\end{tla}
\begin{tlatex}
\@x{ {\VARIABLE} ln}%
\@pvspace{8.0pt}%
\@x{ IsSendAppendEntries \.{\defeq}}%
\@x{\@s{8.2}}%
\@y{%
  Enablement conditions on current state.
}%
\@xx{}%
\@x{\@s{8.2} \.{\land} IsEvent ( ln ,\,\@w{sndAE} )}%
\@x{\@s{8.2} \.{\land} commitIndex [ ln . snd ] \.{=} ln . commit\_idx}%
\@x{\@s{8.2} \.{\land} \.{\dots}}%
\@x{\@s{8.2}}%
\@y{%
  High-level spec actions.
}%
\@xx{}%
\@x{\@s{8.2} \.{\land} AppendEntries ( ln . snd ,\, ln . rcv )}%
\@x{\@s{8.2}}%
\@y{%
  Assertions on successor states.
}%
\@xx{}%
\@x{\@s{8.2} \.{\land} \E\, m \.{\in} Network {\bang} Messages \.{'} \.{:}}%
\@x{\@s{12.29} \.{\land} Network {\bang} OneMoreMessage ( m )}%
\@x{\@s{12.29} \.{\land} IsAppendEntriesRequest ( m ,\, ln )}%
\@x{\@s{8.2} \.{\land} \.{\dots}}%
\@pvspace{8.0pt}%
\@x{ IsRcvAppendEntries \.{\defeq}}%
\@x{\@s{8.2}}%
\@y{%
  Enablement conditions on current state.
}%
\@xx{}%
 \@x{\@s{8.2} \.{\land} IsEvent ( ln ,\,\@w{recvAE} )\@s{4.1}
 \.{\land}\@s{4.1} \.{\dots}}%
\@x{\@s{8.2}}%
\@y{%
  High-level spec actions.
}%
\@xx{}%
 \@x{\@s{8.2} \.{\land} \E\, m \.{\in} Network {\bang} MessagesToFrom ( ln .
 rcv ,\, ln . snd ) \.{:}}%
\@x{\@s{12.29} IsAppendEntriesRequest ( m ,\, ln ) \.{\land}}%
 \@x{\@s{12.29} \.{\lor} HandleAppendEntriesReq ( ln . rcv ,\, ln . snd ,\, m
 )}%
\@x{\@s{12.29}}%
\@y{%
  Impl optimization: Piggyback term on AppendEntries.
}%
\@xx{}%
\@x{\@s{12.29} \.{\lor} UpdateTerm ( ln . rcv ,\, ln . snd ,\, m ) \.{\cdot}}%
\@x{\@s{32.8} HandleAppendEntriesReq ( ln . rcv ,\, ln . snd ,\, m )}%
\@x{\@s{12.29} \.{\lor} \.{\dots}}%
\@x{\@s{8.2}}%
\@y{%
  Assertions on successor states.
}%
\@xx{}%
\@x{\@s{8.2} \.{\land} \.{\dots}}%
\@pvspace{8.0pt}%
\@x{ IsSendAppendEntriesResponse \.{\defeq}}%
\@x{\@s{8.2}}%
\@y{%
  Enablement conditions on current and successor state.
}%
\@xx{}%
 \@x{\@s{8.2} \.{\land} IsEvent ( ln ,\,\@w{sndAER} )\@s{4.1}
 \.{\land}\@s{4.1} \.{\dots}}%
\@x{\@s{8.2}}%
\@y{%
  High-level spec actions.
}%
\@xx{}%
\@x{\@s{8.2} \.{\land} {\UNCHANGED} vars}%
\@pvspace{8.0pt}%
\@x{ IsFault \.{\defeq} \E\, s \.{\in} MultiPowerset ( network ) \.{:}}%
 \@x{\@s{16.4} network \.{'} \.{=} s \.{\land} {\UNCHANGED}
 AllVarsExceptNetwork}%
\@pvspace{8.0pt}%
 \@x{ Spec \.{\defeq} Init \.{\land} {\Box} [ IsFault \.{\cdot} Next ]_{
 {\langle} vars ,\, ln {\rangle}}}%
\end{tlatex}

%% file: 7-results.tex
\section{Results}\label{sec:results}

\renewcommand{\arraystretch}{1.2}
\begin{table}
	\small
	\centering
	\caption{Scale of specifications and state coverage.}\label{tab:spec_sizes}
	\begin{tabular}{clllll}
		\toprule
		& & & &\multicolumn{2}{c}{ Approx states } \\
		& Item & LoC & Vars & /min & Total \\
		\midrule
		\multirow{7}{*}{\rotatebox{90}{Consensus}} & Specification & 1134 & 13 \\
		& Model Checking & 158 & & \oom{6} & \oom{8} \\
		& Simulation & 69 & & \oom{6} & \oom{8}\\
		& Trace Validation & 369 & \\
		& Implementation & 2174 & 25 \\
		& Functional Tests & 2579 & & \oom{5} & \oom{3}\\
		& End-to-end Tests & 2815 & & \oom{3} & \oom{4} \\
		\midrule
		\multirow{4}{*}{\rotatebox{90}{Consistency \hspace{0.07cm}}} & Specification & 375 & 2 & \\
		& Model Checking & 70 & & \oom{6} & \oom{5} \\
		& Simulation & 0 & & \oom{5} & \oom{3}\\
		& Trace Validation & 111 & \\
		& Functional Tests & 123 & \\
		\bottomrule
		\multicolumn{6}{l}{\small All numbers measured on an Azure DC8s v3 VM.} \\
	\end{tabular}
\end{table}

This section summarizes the core results of our efforts, focusing on state coverage as well as the bugs found.
\autoref{tab:spec_sizes} compares the sizes of the \tlaplus{} specs against the implementation and test infra, to give a sense of scale and illustrate the level of detail necessary to execute trace validation.
We find that verification of our \tlaplus{} specs is an extremely efficient way to achieve state coverage.
Comparing state exploration between implementation and specs is straightforward, because traces can also be collected in end-to-end tests, and one log line is largely equivalent to a spec action.
It is immediately apparent that verification of our consensus spec explores orders of magnitude more states at a higher rate than implementation testing.
While code size is not a direct measure of cost, the overall size of the spec and its models are not out of proportion compared to the tests.
Importantly, the consistency spec required very little infrastructure to verify, and to validate traces against the implementation.
The cost of writing formal documentation of the log's consistency guarantee was thus low, and validating it and keeping it in sync with the implementation is equally affordable.

\autoref{tab:bugs_found} lists the most serious bugs found in our consensus protocol as part of the verification work.
These bugs affected both the safety and the liveness of CCF, and were uncovered at several stages of the process by each tool in our verification wardrobe.
In the rest of this section we describe the bugs which were found and corrected during this process, explaining how they were uncovered by our combination of smart casual and classical testing methods.
All these bugs were fixed before they affected any end-users or resulted in customer bug reports.
Thus, although this formalization work took place after the system was initially developed and deployed, it still provided a core benefit of spec-driven development; the identification and removal of critical bugs before they impact production, proving that it is never too late to benefit from smart casual verification.

\begin{table*}
	\caption{Bugs found in CCF's consensus protocol before they could impact production.}\label{tab:bugs_found}
	\small
		\begin{tabular}{llp{10.5cm}}
			\toprule
			Name & Violation & High-Level Description (Issues) \\
			\midrule
			Incorrect election quorum tally & Safety & Quorum was tallied against union of active configurations, rather than against each individual active configuration. (\ghissue{3837}, \ghissue{3948}, \ghissue{4018}) \\
			Commit advance for previous term & Safety & Leaders could advance commit for historical terms without extending log in the current term. (\ghissue{3828}, \ghissue{3950}, \ghissue{3971}, \ghissue{5674}) \\
			Commit advance on AE-NACK & Safety & Variable reuse could cause the leader's commit index to advance when receiving an AE-NACK (\ghissue{5324}, \ghissue{5325}) \\
			Truncation from early AE & Safety & Followers could roll back committed entries, after a sequence triggered by stale AE-NACK messages. (\ghissue{5927}, \ghissue{5991}, \ghissue{6016}) \\
			Inaccurate AE-ACK & Safety & AE-ACK could report an index beyond the end index of AE received, despite the suffix potentially being incompatible. (\ghissue{6001}, \ghissue{6016}) \\
			Premature node retirement & Liveness & Nodes could stop participating in consensus too early during retirement, leading to diminished fault tolerance. (\ghissue{5919}, \ghissue{5973}) \\
			\bottomrule
		\end{tabular}
\end{table*}

\textbf{Incorrect election quorum tally}\label{results/bugs/incorrect_election_quorum_tally} 
48 hours of exhaustive model checking of the consensus spec on a 128 core machine revealed that CCF was tallying election quorums against the union of active configurations (the current configuration plus any pending reconfigurations), rather than against each individual active configuration (compare \S\ref{subsec:ccf/reconfiguration}).
The initial implementation had correctly used the majority in each term, as described in \cite[\S5.2]{Ongaro14}, but had not been updated appropriately when reconfiguration was implemented.
This meant that a node could be elected leader in a term without having a quorum in one of the active configurations, potentially allowing two leaders being elected in the same term, violating a core safety property.
The issue was reproduced with functional and end-to-end testing, a fix was applied, and resolved the problem in both tests and model checking.
This was the first bug identified by TLC and motivated our further investment in \tlaplus.

\textbf{Commit advance for previous term}\label{results/bugs/commit_advance_for_previous_term} 
While assessing the work required to align the implementation and the spec, we discovered that the implementation omitted a check described in Raft.
Our implementation allowed a leader to advance its commit index based solely on receiving a quorum of AE-ACKs confirming a given log entry, and missed the additional requirement that this entry must have been appended by the current leader.
This restriction is fully explained in \cite[\S5.4.2]{Ongaro14}, along with the corresponding risk to safety, but had been accidentally omitted in the implementation.
We added a scenario test based on \cite[Fig. 9]{Ongaro14} to confirm that the implementation was faulty.
An initial fix emptied the node's set of indices eligible for commit (because they are signature transactions) when becoming a leader.
This fix passed all existing and amended tests and was thus integrated into the codebase.
Our work on trace validation, several months later, required us to revise the spec to represent committable indices accurately. Subsequent simulation revealed a safety violation caused by the initial fix; the \emph{fix} broke an implicit property that committable indices contains \emph{all} signatures.
A second fix was implemented and tested as well as verified with TLC.
This bug illustrated that even deterministic testing (compare \S\ref{subsec:trace/existing}) is insufficient to guarantee the correctness of changes.
Moreover, it confirmed that trace validation is effective at guiding the alignment of the spec and the implementation to enable verification.

\textbf{Commit advance on AE-NACK}\label{results/bugs/commit_advance_on_aenack} %
Trace validation discovered that the spec defined a leader's \textit{matchIndex} to remain unchanged after receiving a follower's AE-NACK, whereas the implementation allowed it to decrease.
This difference was due to an aggressive implementation of an optimization proposed in \cite[\S5.3, last paragraph]{Ongaro14}. %
After a single LoC change to align the spec with the implementation, subsequent simulation found a 34-state counterexample violating one of the spec's main correctness properties. %
This counterexample was manually translated into a 150 LoC functional test,
confirming that the implementation could incorrectly advance its commit index.
We also noted that \cite[fig.~2, p.~4]{Ongaro14} implicitly states that \textit{matchIndex} should never decrease, except after a leader election.
Adding this property to the spec allowed model checking to find a shorter counterexample. %
The combination of functional testing and model checking allowed us to fix this bug quickly and confidently.

\textbf{Truncation from early AE}\label{results/bugs/truncation_from_early_ae}
Once a subset of our initial scenarios passed trace validation, investigating why the remaining scenarios failed trace validation uncovered a safety violation.
Log entries necessary to the persistence of committed transactions could be rolled back by a follower.
This bug was introduced by an optimization, and existed in the implementation for some time.%
When the suffix of a follower's log is incompatible with that of the leader, there is a need to find the last agreement point.
The Raft paper describes an iterative reverse search of the sequence numbers, but we instead implemented a suggested optimization to skip entire terms of divergence.
CCF thus finds an agreement point after a sequence of round trips bounded by the number of divergent terms, rather than sequence numbers.
We implemented this optimization by changing the semantics of the {AE-NACK} message, in which followers now include a safe best-estimate of an agreement point communicated using existing fields in the {AE-NACK} message.

Because the leader cannot distinguish these estimate messages from stale {AE-NACK}s emitted in previous terms, it may respond with an {AE} starting before the end of the follower's log.
This, coupled with an insufficiently defensive code path in the follower's code, would cause the {AE} to be treated as a conflicting suffix, and trigger a roll back preceding the application of the {AE}, potentially violating Leader Completeness (defined in \cite[Figure 3]{Ongaro14}).
The fix proved simple: rather than rolling back optimistically on an {AE} in a new term, the follower should only do so on true conflicts.

Notably, this bug was triggered by existing functional tests, producing output that did not pass trace validation, but the tests' assertions were not strong enough. The reproduction scenario we wrote produced a trace 305 events long at the point the bug manifests itself.
This bug showed the important benefit of trace validation to check the invariants in every state, compared to the traditional approach of manually inserting assertions in scenarios at specific points.

\textbf{Inaccurate AE-ACK}\label{results/bugs/inaccurate_aeack}
Fixing this bug directly led to the discovery of the AE-ACK index issue.
Because CCF uses unidirectional messages rather than RPCs (\S\ref{subsec:ccf/consensus}), the code responsible for responding to messages sends values from local state where possible, rather than values specific to the message they responded to.
The AE-ACK handler did so for the \textsc{LastIndex} field, which is the index of the last entry in the appended entries, without correctly checking that the log was also compatible.
This was discovered while conducting trace validation on the previous issue, and a specific scenario was added to test it in isolation.
Here too, the fix was simple, and involved constraining the \textsc{LastIndex} in AE-ACKs to the last index contained within the received AE.

\textbf{Premature node retirement}\label{results/bugs/premature_node_retirement}
Network configuration in CCF is stored in a map (compare \S\ref{subsec:ccf/reconfiguration}).
Adding or removing nodes happens through updates to this map, which produce write sets that are also replicated and handled like any other transaction, and are therefore part of the totally ordered log.
The consensus logic is notified through hooks, which can be called when a transaction is ordered and/or committed.
As a result, the reconfiguration logic is not isolated from the consensus code, and was initially mocked with low fidelity in the scenario driver, and correspondingly simplified in the \tlaplus{} spec.
This was known to be a significant discrepancy between the implementation and the spec, which we decided to address by making the driver more realistic, and by improving the scenario coverage used for trace validation.
As the spec was aligned, simulation produced counterexamples where a reconfiguration would leave the CCF network permanently unable to make progress; a retiring node stopped responding before all future leaders were aware of its retirement.
We translated the counterexamples into new functional tests to reproduce the issue.
A fix, leveraging an existing mechanism to shut down retired nodes safely, was proposed, verified, and implemented together.

\textbf{Non-linearizability of read-only transactions}
Thus far we have described bugs founds in our consensus protocol.
In this last example we describe how we identify an ambiguity in CCF's docs, aided by the consistency spec.
Recall that linearizability requires that transaction execution is consistent with the real-time ordering of client requests and responses.
Recall that \textsc{ObservedRoInv} (Listings~\ref{lst:historyinv}) specifies that any committed read-only transaction must observe any previously committed read-write transactions.
Model checking found a 12-step counterexample to \textsc{ObservedRoInv} in four seconds.
A read-only transaction is handled by an old, yet active leader that has not added a read-write transaction to its log since the new leader was elected.
This is rare in practice, a leader has to be falsely replaced when it is still active.
The leader's logs must be identical, then, in the short window of time before the old leader retires, it needs to handle a read-only transaction.\footnote{The counterexample can be found and interactively explored online~\cite{ccf-6167}.}
Currently, we have no plans to change this behavior, as serializability for read-only transactions is sufficient for most applications, however, we hope that our consistency spec will help us to more clearly communicate this guarantee to developers.

%% file: 8-lessons.tex
\section{Lessons Learned}\label{sec:lessons}

We were surprised that, despite our extensive testing, \emph{many bugs were first spotted during the development of the spec and subsequent alignment with the implementation.}
The development and refinement of the consensus and consistency specs forced us to think deeply about our protocol and its invariants.
During the process of spec development, the implementation was very closely scrutinized (with the target invariants in mind) and many bugs were first identified.
In some cases, someone would become suspicious of a particular part of the code but would be unable to confirm that the current behavior led to a violation.
This was particularly true for situations including multiple reconfigurations and failure handling (see Incorrect election quorum tally \S\ref{results/bugs/incorrect_election_quorum_tally} and Commit advance for previous term \S\ref{results/bugs/commit_advance_for_previous_term}), where counterexamples require many steps and were no longer feasible to work through on a whiteboard.
This is where verification was invaluable, as it allowed us to quickly check the behavior against the expected invariants.

While finding implementation bugs by gradually and manually aligning a formal spec with its implementation is possible, we found that \emph{trace validation is a more systematic and efficient approach}. 
Prior to our trace validation efforts, which began in Spring 2023, we were not confident that our consensus spec matched the implementation.
Different team members worked on the spec and the implementation, and as such, they reflected different understandings of how the consensus worked.
Moreover, we corrected safety violations present only in the spec, while the implementation contained bugs that spec verification could not find.
Trace validation, and its inclusion in our CI pipeline, proved to be a turning point in our verification efforts, as it finally allowed us to systematically identify and fix these discrepancies.

Our results show that \emph{that software verification is beneficial even for systems that have already been ``proven in production''}.
During the time that the incorrect election quorum tally (\S\ref{results/bugs/incorrect_election_quorum_tally}) was present in CCF, the operators added and removed nodes one at time.
This meant that the bug did not lead to a safety violation and thus remained undiscovered, but could have surfaced if the operators had changed their reconfiguration strategy.
Similarly, that the incorrect fix for the commit advance for previous term bug (\S\ref{results/bugs/commit_advance_for_previous_term}) did not lead to a production incident can be attributed to chance.

Despite modern hardware and advances in tooling, we found that \emph{exhaustive model checking, at our level of abstraction, took significant time to complete}.
This led us to limit the state space more than we would have liked.
We could have used a proof system, such as the \tlaplus Proof System, instead of a model checker.
By opting for model checking, we chose to prioritize developer time and accessibility over compute time. %
As interactive theorem provers continue to advance~\cite{coq,paulson90}, along with AI-assisted verification~\cite{Lample22}, this trade-off may no longer be necessary.
However, \emph{simulation proved effective at quickly finding bugs}, especially when combined with action weighting (described in \S\ref{sec:consensus}).

%% file: 9-related.tex
\section{Related Work}\label{sec:related}

Validating the correctness of distributed systems is a widely studied problem with approaches ranging from rigorously verified implementations, \emph{formal verification}, to the many flavors of software testing, \emph{casual verification}, including \emph{smart casual verification} which sits between the two.

Formal verification of distributed systems provides the strongest guarantees of correctness, but is often impractical for real-world systems due to the high cost of development and expertise required.
For example, IronFleet~\cite{Hawblitzel17} and Verdi~\cite{Wilcox15} both proved implementations of Raft correct, but, to the best of our knowledge, have not been used outside of an academic setting.
Moreover, they are not easily amenable to systems already implemented in a general-purpose programming language.
PGo\cite{hackettCompilingDistributedSystem2023} follows a related approach in which one could prove the correctness of a \tlaplus{} spec, and then extract a Go implementation using their PGo compiler. Again, an approach that works best for greenfield projects.

On the other hand, there is a broad spectrum of approaches to testing distributed systems (casual verification)~\cite{jepsen,Brooker20,Meng23}, which tend to follow the same pattern: (i) orchestrate the creation of one or more configurations of the system, (ii) schedule workloads, and (iii) inject faults, such as network partitions, node failures, clock skew etc.
As the extent of the system and its dependencies being orchestrated increases, it becomes more difficult to maintain determinism and repeatability.
Test times for equivalent scenarios also tend to grow longer, and the likelihood of spurious failures goes up.

\summary{\tlaplus and consensus} 
CCF's consensus spec is the latest addition in a recent tradition of modeling consensus in \tlaplus{}.
This began with Lamport's description of Paxos~\cite{Lamport98} in \tlaplus{}, as a refinement of higher-level specs~\cite{paxos-tla}.
This was followed by the formalization in \tlaplus{} of other consensus algorithms~\cite{Yin20,tendermint-tla,Schultz22} including Paxos variants~\cite{Lamport11,byzpaxos-tla,mpaxos-tla,Gafni00,Lamport06,Howard17} and Raft~\cite{raft-tla, Schultz24}.

Tasiran et al.~\cite{tasiran2002} were the first to extract and validate traces obtained from a hardware simulator against a \tlaplus{} spec, demonstrating the practical applicability of trace validation.
The adoption of \tlaplus{} among distributed system practitioners, spurred by Newcombe et al.~\cite{Newcombe2015}, and the formalization of trace validation as a refinement check by Pressler~\cite{Pressler2020}, caused trace validation to be applied to real-world distributed systems.
For instance, Davis et al.~\cite{davis_extreme_2020} applied the technique to MongoDB, discovering a non-trivial implementation bug.
However, they faced challenges in consistently logging implementation state, and aligning different grains of atomicity, which we attribute to them not leveraging \tlaplus's non-determinism to infer implementation state, and action composition to align atomicity. Niu et al.~\cite{niuVerifyingZookeeperBased2022} also validated traces of Zookeeper, ensuring that its implementation corresponds to its spec.
Similarly, Wang et al.~\cite{wangModelCheckingGuided2023} revealed several implementation bugs by replaying \tlaplus{} behaviors against instrumented implementations. Likewise, SandTable is capable of replaying behaviors but, additionally, provides a generic testing framework for distributed systems~\cite{sandTableScalableDistributed2024}. SandTable intercepts network communication at the POSIX layer, making it incompatible with systems like CCF that encrypt communications at the application layer. Furthermore, Wang et al. and SandTable serve as examples of the challenges of aligning the grains of atomicity, illustrated by the authors identifying two bugs in Ongaro's well-established Raft~\cite{raft-tla} spec. We contend that these are, in fact, common \tlaplus{} modeling patterns and can be handled with action composition.  Nevertheless, all efforts found non-trivial bugs in real-world systems by comparing implementation traces to high-level \tlaplus{} behaviors; a testament to the effectiveness of this approach.

More pragmatic verification efforts are not limited to TLA+; Amazon's S3 ShardStore service was recently augmented with \textit{lightweight verification}~\cite{bornholt21} using reference models written in Rust, which are simplified instantiations of program components that can be used to track program state under different input conditions. Like CCF, the primary goals are usability, and the ability to ensure correctness as both the implementation and the spec evolve over time.
Unlike the CCF approach however, state exploration is limited to what the test harness is able to reach.

%% file: 10-conclusion.tex
\section{Conclusion}\label{sec:conclusion}

This paper details our journey with smart casual verification of the distributed protocols in CCF using \tlaplus{}. 
Our experience demonstrates that \tlaplus{} can be used in industrial settings to verify extensive and nuanced distributed protocols, and that the verification process can be integrated into the development workflow of a production codebase. 
We have seen that \tlaplus{} can be effectively utilized to find bugs in both the design and implementation of these protocols and to communicate understanding of complex and subtle distributed systems like CCF.